\documentclass[aps,pra,amsmath,amssymb,amsfonts,superscriptaddress,floatfix,nofootinbib]{revtex4} 
\pdfoutput=1
\usepackage{amsmath,amsthm,amsfonts,amssymb}
\usepackage{graphicx}

\usepackage{color}
\usepackage{hyperref}

\usepackage{algorithm}
\usepackage[capitalize,nameinlink]{cleveref}

\newcommand{\be}{\begin{equation}}
\newcommand{\ee}{\end{equation}}

\newcommand{\hc}{{\rm h.c.}}
\newcommand{\Hq}{H_{\rm quad}}

\begin{document}

\title{Improving Perturbation Theory with the Sum-of-squares II: Large Density-Density Terms}
\author{Matthew B.~Hastings}
\begin{abstract}
In Ref.~\cite{hastings2024improving}, a method was given for self-consistently generating sum-of-squares decompositions of quartic fermionic Hamiltonians.  Perturbation theory was used to generate a useful choice of cubic operators in this sum-of-squares.  On a range of model problems, this method, which is only a fragment of degree-six sum-of-squares, was able to outperform the full degree-four sum-of-squares in both speed and accuracy.
Unfortunately for applications, many problems in chemistry have strong density-density interaction terms, as well as moderately strong density-dependent hopping and spin-spin interaction terms, limiting the power of the perturbative choice of the cubic operators.  Here we propose a method for generating these decompositions in the presence of these strong interaction terms, hopefully extending the range of applicability of this method.
\end{abstract}
\maketitle

\section{Introduction}
The reduced density matrix method (RDM)\cite{coleman1963structure,erdahl1978representability,
percus1978role,mazziotti2001uncertainty,Nakata_2001,Maz12,klyachko2006quantum} is a method used in quantum chemistry to prove lower bounds on the energy of quantum Hamiltonians.  It is an instance of the sum-of-squares hierarchy for quantum Hamiltonians.  While the time required is likely to be polynomial at each level of the hierarchy (at least, it requires solving a semi-definite program (SDP) of polynomial size at each level), the degree of the polynomial increases with the level, so that one of the most common forms is the lowest nontrivial level, the degree-4 sum-of-squares.  The degree-4 sum-of-squares is also called the 2RDM (2 reduced-density-matrix), as it tracks two-particle correlations.
This 2RDM has some issues, however; in particular, it was shown\cite{hastings2022perturbation} that it does not reproduce second-order perturbation theory for fermionic quantum Hamiltonians with quartic interactions, though degree-6 sum-of-squares does.
In Ref. \cite{hastings2022perturbation}, it was shown also that one could reproduce second- and third-order perturbation using a certain simple sum-of-squares decomposition involving only squares of operators which are either linear or cubic in the fermion creation and annihilation operators; those cubic terms were similar to those used in \cite{hastings2022optimizing} for the SYK model, but here applied to a weakly interacting setting.
In \cite{hastings2024improving}, it was shown that one could rapidly find such a decomposition using a self-consistent method; the advantage of this method is that it does not require solving a semi-definite program\footnote{To be more precise, in a generalized algorithm of Ref.~\cite{hastings2024improving}, one does solve a semi-definite program, but the semi-definite program is the same kind as one solves when optimizing a single-particle fermionic Hamiltonian, where one optimizes a single-particle density matrix subject to all eigenvalues being in the interval $[0,1]$.  Hence, this semi-definite program reduces to matrix diagonalization, and in this case it is of a matrix of size proportional to the number of fermion modes.  We discuss a similar generalized algorithm applied to the method here at the end of this paper.} while it still yields strict lower bounds on the energy.

Further, this self-consistent sum-of-squares method significantly outperformed the 2RDM in both speed and accuracy for a range of model Hamiltonians, and also outperformed third-order perturbation theory in accuracy.  These model Hamiltonian involved a quadratic part plus a randomly chosen quartic part, with all quartic terms chosen independently and of the same magnitude.  We emphasize that while we used ideas from perturbation theory to guide the choice of the terms in the decomposition, the method is non-perturbative, yielding lower bounds on the energy and handling problems which perturbation theory could not handle.  

There was, however, a major barrier in applying this approach to quantum chemistry.  The problem is that relevant quantum chemistry Hamiltonians often have certain strong terms in the Hamiltonian, with the largest corresponding to density-density interactions.  These terms are large enough that the decomposition we chose did not accurately treat them.  In this paper, we give an approach to dealing with these terms and also certain other large terms.  Indeed, we provide more than one approach, but we suggest one approach in particular based on the low degree of terms needed to obtain a certain closure property for the self-consistent method, explained below.

This paper may be seen as a sequel to \cite{hastings2022optimizing,hastings2022perturbation,hastings2024improving}, and for that reason we omit giving some background which is standard in the field of the sum-of-squares.

\section{Background and Relevant Large Interaction Terms}
We consider fermionic Hamiltonians of the form
\be
H=H_0+\epsilon V
\ee
where initially we take
$$H_0=\sum_i e_j \psi^\dagger_j \psi_j,$$
for some scalars $e_j>0$, where $\psi^\dagger_j,\psi_j$ are fermionic creation and annihilation operators, with a total of $n$ fermionic modes, where $\epsilon$ is a scalar, and
where
$V$ is quartic in the fermionic creation and annihilation operators and $V$ is normal ordered.
In Ref.~\cite{hastings2024improving}, this general form was considered, but then later spin and number conservation were used for speed in a numerical implementation of the self-consistent method.  Here, we will not use spin and number conservation.

Further, our choice of $H_0$ assumes that all $e_j$ are positive.  Doing this may require performing a particle-hole conjugation on some of the fermioni modes.

\subsection{Self-Consistent Sum-of-Squares}
The self-consistent sum-of-squares method of Ref.~\cite{hastings2024improving} finds a sum-of-squares decomposition of
$H$ as
\be
\label{Hdecomp}
H=
 \sum_i e'_i \Bigl( \psi'_i+\tau_i \Bigr)^\dagger \Bigl( \psi'_i + \tau_i  \Bigr) +\sum_i e'_i \tau_i \tau_i^\dagger+\lambda,
 \ee
 proving that the lowest eigenvalue of $H$ is at least some scalar $\lambda$.  Here, the $e'_i$ are some new non-negative scalars, and the $\psi'_i$ are some unitary rotations of the $\psi_i$, and the $\tau_i$ are odd polynomials of degree at most three in the fermionic operators.
This use of $\tau$ variables is similar to Ref.~\cite{hastings2022optimizing}, but the self-consistent method gives a way to find these variables.
Note that the sum-of-squares decompositions we consider are sum of operators of the form $O^\dagger O$ for some (possibly non-Hermitian) operator $O$.

To find this decomposition, we pick some other Hamiltonian, $H'$, called a trial Hamiltonian, which has quadratic part $H'_0=\sum_i e'_i (\psi'_i)^\dagger \psi'_i$.  We have some prescription for constructing a set of $\tau_i$ for a given trial Hamiltonian.
Then, from this set of $\tau$, we construct a sum-of-squares decomposition such that
 \be
 \label{sosd}
 \sum_i e'_i \Bigl( \psi'_i+\tau_i \Bigr)^\dagger \Bigl( \psi'_i + \tau_i  \Bigr) +\sum_i e'_i \tau_i \tau_i^\dagger
=H' + \sum_i e_i \{ \tau_i,\tau_i^\dagger\}.
\ee
Let
\be
\sum_i e'_i \{ \tau_i,\tau_i^\dagger\} = W -\lambda,
\ee
for some scalar $\lambda$,
where $W$ is at most quartic in the creation and annihilation operators, and where $W$ has vanishing scalar part when written in normal-ordered form.
Then, from this trial Hamiltonian, we define an effective Hamiltonian $H_{eff}=H'+W$ so that \cref{sosd} gives a sum-of-squares decomposition of $H_{eff}$.  If we can adjust the trial Hamiltonian $H'$ so that the effective Hamilton $H_{eff}$ is equal to $H$, then we have found the desired sum-of-squares decomposition of $H$.  Since the operators $\tau_i$ constructed by this prescription are $O(\epsilon)$, the resulting $W$ is $O(\epsilon^2)$ and so for small $\epsilon$ one may find such a trial Hamiltonian by a simple iterative procedure: start with $H'=H$, then compute the effective Hamiltonian, adjust the trial Hamiltonian by subtracting off the difference between $H_{eff}$ and $H$, recompute the effective Hamiltonian, and so on, until convergence.

This method has two important properties.  First,
the error in ground state energy (i.e., the difference between the true ground state energy and the bound from this decomposition) is $O(\epsilon^4)$ because\cite{hastings2024improving} each term in the sum-of-squares is a square of a term that annihilates the ground state up to error $O(\epsilon^2)$.  We shall refer to this property that each term annihilates the ground state up to error $O(\epsilon^2)$ as the ``error property".  The second property it has is that the effective Hamiltonian is equal to the trial Hamiltonian up to terms of order $O(\epsilon^2)$; this allows the iterative solution of the equation $H_{eff}=H$, and also guarantees that any decomposition that obeys the error property for $H'$ will also obey it for $H_{eff}$.  We call this property the ``closeness property" as $H_{eff}$ is ``close" to $H'$.  A final property is the effective Hamiltonian is of the same form as the trial Hamiltonians we consider; i.e., in this case, it is an arbitrary quartic Hamiltonian.  We call this the ``closure property".

We emphasize that the use of perturbation theory here is to construct a good choice of operators $\tau_i$.  A more general procedure optimizing over these possible operators might be considered, solving the sixth order sum-of-squares, but this would be more numerically expensive.  We have termed the procedure ``self-consistent", but there really are two ingredients in this procedure: first, a particular decomposition using squares of operators which are fermion-odd, and, second, a self-consistent method for solving this using trial and effective Hamiltonians.

This procedure lower bounds the ground state energy and, as shown\cite{hastings2024improving}, the bound is within $O(\epsilon^4)$ of the true ground state energy.  Unfortunately, while the method does seem to be numerically more accurate and faster than the 2RDM for a class of problems where all interactions terms are sufficiently small (but still outside the regime of perturbation theory)\cite{hastings2024improving}, it does not apply well to many quantum chemistry problems in this form because certain interaction terms in the Hamiltonian are not small.
We discuss these in \cref{largeterm}.

\emph{Before continuing, we note that from now on we will drop many of the primes that appear as superscripts when writing the sum-of-squares for simplicity of presentation.}  The self-consistent method gives a sum-of-squares which approximates $H'$ (i.e., it obeys the closeness property), and adjusting $H'$ until finally the decomposition is equal to $H$.  Then, operators $\psi'$ are used as annihilation operators as they are eigenoperators of $H'_0$.  However, to avoid writing primes excessively, we will instead, whenever writing a sum-of-squares decomposition, write it as some decomposition which approximates a Hamiltonian $H$, without primes.  Thus, we will let $H$ be our trial Hamiltonian, and adjust $H$ until the sum-of-squares decomposition gives some other given ``target Hamiltonian".

\subsection{The Relevant Large Interaction Terms}
\label{largeterm}

The particular large terms seem to be density-density interaction, density-dependent hopping, spin-spin interaction, and singlet hopping.  The strongest of these is density-density interaction,
which is of the form $$\sum_{i<j} U_{ij} n_i n_j,$$ where $n_i=\psi^\dagger_i \psi_i$ is the number operator.  This term may have either sign.  Of course, for actual, physical electrons, the interaction is repulsive, but since we have performed a particle-hole conjugation on some of the modes, the sign may be positive or negative.  Indeed, we expect $U_{i,j}$ to be positive iff we have performed particle-hole conjugation on \emph{both} mode $i$ and mode $j$ or \emph{neither} mode $i$ nor mode $j$.

So, given the strength of density-density interaction, we would not like to include it in $\epsilon V$ which is $O(\epsilon)$ but we would like to modify the possible $H_0$ that we consider  so that density-density interaction can be included there.  Remark: since we will see that it is more complicated to include density-density interaction in $H_0$, one may wish to only include the strongest density-density terms in $H_0$ and include the weaker density-density terms in $\epsilon V$.

Density-dependent hopping is a linear combination of terms of the form
$$n_i (\psi^\dagger_j \psi_k + \hc).$$  In this paper, we will primarily be concerned with handling density-density interaction; that is, with finding a decomposition whose error is still $O(\epsilon^4)$ even if the density-density terms themselves are not $O(\epsilon)$, and where the deomposition can be computed with effort not much more than that used when all terms are $O(\epsilon)$.
As a side effect we will also be able to handle density-depending hopping in a similar way (at least in the case that the hopping is of the given form after particle-hole conjugation of modes, rather than being $n_i (\psi^\dagger_j \psi^\dagger_k+\hc)$  after conjugation), and we will be able to handle spin-spin interactions ``for free"; that is, the machinery that handles density-density interactions will handle them as well so that they may be in $H_0$ also.  However, one may wish to include these spin-spin terms in $\epsilon V$ for simplicity as the terms are smaller than density-density.

Since we are concerned with applications to chemistry Hamiltonians, which have spin-rotation symmetry, we emphasize that in that application there is additionally a spin index, so each mode is labelled by a pair $i,\sigma$.  Then, the density-density interactions are of the form
$\sum_{i<j} \sum_{\sigma,\tau\in\{\uparrow,\downarrow\}} U_{ij} n_{i,\sigma} n_{j,\tau}.$  However, each such spin-rotation symmetric interaction is equal to the sum of four different density-density interactions above, between different spin orbitals $(i,\uparrow)$, $(i,\downarrow)$, $(j,\uparrow)$, $(j,\downarrow)$, so the procedure developed here applies to that case, though of the use of symmetry does simplify some numerics.

Similarly, the density-dependent hopping is a linear combination of terms of the form  $\sum_{\sigma,\tau\in \{\uparrow,\downarrow\}} n_{i,\sigma} (\psi^\dagger_{j,\tau} \psi_{k,\tau} + \hc),$ which again is of the form of terms considered here.

Finally, the spin-spin interaction is a linear combination of terms of the form
$\sum_{\mu,\nu,\rho,\sigma} \psi^\dagger_{i,\mu} P_{\mu\nu} \psi_{i,\nu} \psi^\dagger_{j,\rho} P_{\rho\sigma} \psi_{j,\sigma},$
where $P$ is a Pauli matrix, and one sums over $P$ being $X,Y,$ or $Z$.  One sees, then, that if $P$ is the Pauli $Z$ matrix, then this is a sum (with signs) of four density-density interactions between spin-orbitals $(i,\uparrow)$, $(i,\downarrow)$, $(j,\uparrow)$, $(j,\downarrow)$, and so it can be handled by the same procedure as here.  Again, the use of symmetry would be to simplify the calculations.
What about the case where $P$ is Pauli $X$ or $Y$?  These terms are density-density interactions between a set of rotated spin-orbitals; i.e., if we chose a different single-particle basis, we would have these terms.  

Applying a particle-hole conjugation to density-density interaction or spin-spin interaction terms produces a term of the same form, up to possibly a change in sign and possibly producing quadratic terms.  For example, if we apply a particle-hole conjugation to mode $i$ but not to mode $j$, then a term $U_{ij} n_i n_j$ becomes $-U_i n_i n_j+n_j$.
Applying particle-hole conjugation to density-dependent hopping produces a density-dependent hopping if particle-hole conjugation is applied to \emph{both} of the modes, but if it is applied to only one, then it produces density-dependent pair creation, which we will not include, as mentioned above.
We will see then that with the methods here we can handle the density-density and spin-spin interaction terms, regardless of particle-hole conjugation, and also (depending on particle-hole conjugation) density-dependent hopping terms.

We will not be concerned with developing any new method for the singlet hopping term.  First, these terms, while larger than some quartic terms, are relatively small.  Second, the self-consistent sum-of-squares already numerically outperforms the 2RDM for these terms, at least for some model Hamiltonians where we have a quadratic term plus randomly chosen singlet hopping terms in some range of interaction strengths\cite{hastings2024improving}.  Finally, we should not expect an easy way to handle these terms!  The other three types of terms (density-density, density-dependent hopping, and spin-spin) all have the property that they annihilate the the ground state of the quadratic part of the Hamiltonian invariant.  On the other hand, singlet hopping does not in general annihilate this state: singlet hopping is some term (before particle-hole conjugation) $\psi^\dagger_{i,\uparrow} \psi^\dagger_{i,\downarrow} \psi_{j,\uparrow} \psi_{j,\downarrow}+\hc$ and so if we particle-hole conjugate mode $j$, this term creates four fermions.

\section{Changing the $\tau$ Terms}
The first thing needed for any approach to the density-density interaction is a change in the definition of $\tau_i$ as we need  the error property to still hold, so we want $\psi_i+\tau_i$ to annihilate the ground state to $O(\epsilon^2)$.
We will assume a Hamiltonian $H$ of the form
\be
\label{intform}H=H_0+ \epsilon V,
\ee
where now we allow
\be
H_0=\Hq+\sum_{i<j} U_{i,j} n_i n_j,
\ee
with
\be
\Hq=\sum_i e_j \psi^\dagger_j \psi_j
\ee
containing the quadratic parts.  That is, we have moved the density-density terms from $V$ to $H_0$.

Here we are only include density-density terms in $H_0$, and not density-dependent hopping.  We will include those later in the approach we use in \cref{ascscsos}

 We normal order the terms in $V$, meaning that the annihilation operators are to the right of creation operators.   We
$V_{ijkl}$ to denote the coefficient of $\psi_i \psi_j \psi_k \psi_l$ in $V$.  
To denote terms with both creation and annihilation operators, we will
use overlines to denote creation operators, so that
$V_{\overline i \overline j k l}$ is the coefficient of $\psi^\dagger_i \psi^\dagger_j \psi_k \psi_l$, and similarly for terms with other numbers of creation and annihilation operators.
Since $V$ is normal ordered, we write coefficients with an overline (corresponding to creation) to the left of those with no overline (corresponding to annihilation), and we let $V_{abcd}$ be totally anti-symmetric in any set of indices which all have an overline or all do not have an overline, so that
\begin{align}
V=&\Bigl(\sum_{i<j<k<l} V_{ijkl} \psi_i \psi_j \psi_kj \psi_l+\hc\Bigr) + \Bigl(\sum_{i} \sum_{j<k<l} V_{\overline i j k l} \psi^\dagger_i \psi_j \psi_k \psi_l+\hc\Bigr)
\\ \nonumber &+ \sum_{i>j} \sum_{k<l} V_{\overline i \overline j k l} \psi^\dagger_i \psi^\dagger_j \psi_k \psi_l.
\end{align}

The operators $\tau_i$ were previously defined by
\begin{align}
\label{taudef}
\tau_i=\epsilon \sum_{j<k<l} T^{\overline i}_{\overline j\overline k \overline l} \psi^\dagger_j \psi^\dagger_k \psi^\dagger_l+
\epsilon \sum_{j<k} \sum_l T^{\overline i}_{\overline j\overline k l} \psi^\dagger_j \psi^\dagger_k \psi_l+
\epsilon \sum_{j} \sum_{k<l} T^{\overline i}_{\overline j k l} \psi^\dagger_j \psi_k \psi_l,
  \end{align}
where the scalar $T$ is given by
\begin{align}
\label{olddenom}
T^{\overline i}_{\overline j\overline k \overline l}=\frac{1}{e_i+e_j+e_k+e_l} V_{\overline i \overline j \overline k \overline l},
\\ \nonumber
T^{\overline i}_{\overline j\overline k  l}=\frac{1}{e_i+e_j+e_k} V_{\overline i \overline j \overline k l},
\\ \nonumber
T^{\overline i}_{\overline j k  l}=\frac{1}{2}\frac{1}{e_i+e_j} V_{\overline i \overline j k l},
\end{align}
and where $e_i$ are the coefficients of the quadratic terms in the Hamiltonian $H$ so that $\Hq=\sum_i e_i \psi^\dagger_i \psi_i$.
Then, $\psi_i+\tau_i$ has the error property so long as all terms in the quartic interaction are $O(\epsilon)$.  However, if the density-density interaction is of order unity, we should change the energy denominators to
\begin{align}
\label{newdenom}
T^{\overline i}_{\overline j\overline k \overline l}=\frac{1}{e_{i,j,k,l}} V_{\overline i \overline j \overline k \overline l},
\\ \nonumber
T^{\overline i}_{\overline j\overline k  l}=\frac{1}{e_i+e_j+e_k} V_{\overline i \overline j \overline k l},
\\ \nonumber
T^{\overline i}_{\overline j k  l}=\frac{1}{2}\frac{1}{e_i+e_j} V_{\overline i \overline j k l},
\end{align}
where
$$e_{i,j,k,l}=e_i+e_j+e_k+e_l+U_{i,j}+U_{i,k}+U_{i,l}+U_{j,k}+U_{j,l}+U_{k,l}.$$
The physical interpretation of $e_{i,j,k,l}$ is that it is the expectation value of $H_0$ in a state with modes $i,j,k,l$ occupeid and all others empty.

\section{Possible Approaches}
\label{possapp}
Here we discuss a few possible approaches to handling with the density-density interaction, and give the problems with these approaches.
One may prefer to skip to \cref{ascscsos} on first reading; that section gives a simpler approach which lacks these problems.

First, suppose $U_{ij}\geq 0$ for all $i,j$.  Then, we can write the sum-of-squares decomposition
$$\sum_{i<j} U_{i,j} n_i n_j=\sum_{i<j} U_{ij} (\psi_j \psi_i)^\dagger (\psi_j \psi_i).$$  These terms do not obey the error property  but we can adjust them by $O(\epsilon)$ so that they do, replacing each term instead as
\be
\label{replace}
U_{ij} n_i n_j \rightarrow
U_{ij} \Bigl|\psi_j \psi_i+\sum_{k,l} \frac{V_{k,l}}{e_{i,j,k,l}} \psi^\dagger_k \psi^\dagger_l \Bigr |^2,
\ee
and adding that term in the sum-of-squares decomposition,
where here we adapt the convention that for any operator $O$, the expression $|O|^2$ denotes $O^\dagger O$.
That is, the full sum-of-squares decomposition would be the sum of these terms, plus the previous sum-of-squares decomposition except with
$\tau$ given by \cref{newdenom} rather than by \cref{olddenom}.

Now we show that this obeys the closeness property.  One may explicitly check this.  However, there is a simpler way.  First,
 the
replacement \cref{replace} includes the terms $U_{ij} n_i n_j$, but also includes some additional quartic terms of order $\epsilon$ and $\epsilon^2$.
These new quartic terms of order $\epsilon$ involve either four creation operators or four annihilation operators.
Also, using \cref{newdenom} for $\tau$ gives the same decomposition as \cref{olddenom} for all terms except those involving
four creation or four annihilation operators (i.e., any term with both creation and annihilation operator is the same).  Since \cref{olddenom} leads to an $H_{eff}$ with the correct (up to error $O(\epsilon^2)$) quartic terms  in $V$ , one may see that this new decomposition obeys the closeness property up to, possibly, whether or not $H_{eff}$ has the correct terms involving four creation or four annihilation operators.
The question one must check is whether there is a cancellation: is the change in those terms with four creation or annihilation operators from changing the definition of $\tau$ cancelled by the new quartic terms resulting from \cref{replace}.
However, if it did not correctly give those terms in $H_{eff}$ up to error $O(\epsilon)$, then it would not obey the error property as follows.  By construction, the sum-of-square decomposition obeys the error property \emph{for the ground state of} $H$, and since this decomposition is a sum-of-squares decomposition of $H_{eff}$, if we compute the expectation value of $H_{eff}$ in the ground state of $H$, the result is equal to the ground state energy of $H_{eff}$ up to $O(\epsilon^4)$, and so, since $H$ and $H_{eff}$ are gapped (for small $\epsilon$) the ground state of $H$ is equal to the 
ground state of $H_{eff}$ up to $O(\epsilon^2)$.  However, if we had an error in the terms with four creation or annihilation operators which was of order $\epsilon$, then the ground states would differ by order $\epsilon$.

This approach also has the nice advantage that $H$ and $H_{eff}$ are both quartic.
On the other hand, the downside is the restriction to $U_{ij}$ being all positive.

So, one should consider another approach.

A second possible approach is as follows.  We would like the state with $n_i=0$ for all $i$ to be the ground state of \cref{intform} if $V=0$.  A sufficient condition for this to hold is that $H_0=\Hq+\sum_{i<j} U_{i,j} n_i n_j$ can be written as a sum-of-squares, with each term in the sum of squares being linear in the $n_i$ and using that $(n_i)^2=n_i$.  That is, each term is of the form $(\sum_i a_i n_i)^2$ for some scalars $a_i$ which may be taken real.

Then, replace each term with $(\sum_i a_i \tilde n_i)^2$ where $\tilde n_i=(\tilde \psi_i)^\dagger \tilde \psi_i$ and $\tilde \psi_i=\psi_i+\tau_i$.
Then, each term in this sum-of-squares obeys the error property.  Suppose that one then succeeds in showing that the quartic terms of order $\epsilon$ that results from this replacement are equal to $\epsilon V$ up to $O(\epsilon^2)$.  However, this will still not necessarily give us the closeness property as one may have some extra terms appear of the form such as $\epsilon n_i V_i$ where $V_i$ is some arbitrary quartic, and these terms are also order $\epsilon$.  Further, these terms would prevent us from having the closure property.
Also, even higher order (in $\epsilon$ and $n$) terms will appear.

In the rest of this section, we discuss one way of dealing with these terms such as $\epsilon n_i V_i$.  We only sketch this approach, since in
\cref{ascscsos} we give a completely different approach to handling the density-density terms.  The approach of \cref{ascscsos} will also require dealing with terms like $\epsilon n_i V$, and we give a different way to handle them here, different from what we consider in the rest of this subsection.

First, we may assume that $V_i$, when expressed in terms of operators $\psi^\dagger,\psi$ does not contain terms $\psi_i$ or $(\psi_i)^\dagger$, as any such terms can be combined with $n_i$ to obtain quartic terms.
If $V_i$ does not contain such terms then we expect $\epsilon n_i V_i$ to have expectation value $O(\epsilon^4)$: roughly speaking the leading $\epsilon$ gives one factor of $\epsilon$, the $n_i$ gives a factor of $\epsilon^2$, and the $V_i$ gives another factor of $\epsilon$.

In fact, there is a general way to prove this using the sum-of-squares, or, more precisely, to add such terms while only changing the bound on the ground state energy by $O(\epsilon^4)$.  We can normal order such a term in terms of operators $\psi^\dagger,\psi$ and it has at least one operator $\psi^\dagger$ and at least one operator $\psi$ in this normal-ordered form.  Then, we can write the rightmost $\psi$ as $\tilde \psi$ plus $\epsilon$ times a cubic in $\psi^\dagger$, and we can do something similar to the leftmost $\psi^\dagger$.  If what is left after this is something of the form $\tilde \psi^\dagger \ldots \tilde \psi$, with $\ldots$ denoting some operator, then this, by Cauchy-Schwarz, is bounded by some linear combination of terms $\tilde n_i$, and adding such terms to the Hamiltonian will not change the bound on the ground state energy so long as the coefficient of these terms is sufficiently small.  The lowest order thing that could be left instead is something of the form $\epsilon^2$ times a product of eight annihilation or eight creation operators.  One could then shift $\tau$ by adding in $\epsilon^2$ times a product of $7$ creation operators to produce this term in the Hamiltonian, and it is possible to show that this would change the bound on the ground state energy by $O(\epsilon^4)$, though one would then be left with some additional terms in the Hamiltonian which are $\epsilon^4$ times the anti-commutator of a product of $7$ creation operator with $7$ annihilation operators (and hence this anti-commutator is of degree $12$).

Clearly, this becomes rather complicated, and that is why we prefer the method of the next section.

\section{A Simple Closed Self-Consistent Sum-of-Squares}
\label{ascscsos}
For the method of this section,
we will expand our choice of Hamiltonian $H$ so that
\be
\label{intformnew}H=H_0+ \epsilon V+\sum_{i} \epsilon n_i V_i,
\ee
where $V_i$ is a quartic, like $V$, with $V_i$ not including $\psi_i$ or $\psi_i^\dagger$.

We also now allow $H_0$ to include density-dependent hopping terms so that
\be
H_0=\Hq+\sum_i n_j (\Hq)_i,
\ee
with $\Hq(i)$ being of the form
$\sum_{j,k} \psi^\dagger_j T_{j,k}(i) \psi_k$, for some matrix $T$ depending on $i$.

The term $\sum_{i} \epsilon n_i V_i$ is added for the closure property, as we will see.

\subsection{The Decomposition}
We will decompose $H$ as a sum of terms $H=\sum_a c_a ( |F_a|^2+|\theta_a^\dagger|^2)$, where $c_a$ are non-negative scalars, and each
$F_a$ takes the form
$$F_a=\psi_i + n_j  \sum_{k\neq j} d_k \psi_k+\theta_a,$$
where
$$\theta_a=\tau_i + \mu+\epsilon n_j \nu.$$
Here $i,j$ and $\tau$ depend on $a$.  Here, $\tau_i$ is as in \cref{taudef},\cref{newdenom}.  The scalars $d_k$ depend on $a$ also, and
$\nu$ is a cubic in the fermion operators also depending on $a$.
Finally, $\mu$ is also a
cubic in the fermion operators and also depends on $a$, and we pick $\nu$ so that it commutes with $n_j$. 
The term $\theta_a$ will be $O(\epsilon)$.

The role of the term $\mu$ here is to ``correct" the $\tau_i$ term because of the term $n_j  \sum_{k\neq j} d_k \psi_k$ in $F_a$, so that we still obtain the error property.  The role of the $\nu$ term is to construct terms $\epsilon n_i V_i$ in the Hamiltonian, to obtain the closure property as we will see.

Note that $\psi_i+\tau_i$ obeys the error property by construction, i.e., it annihilates the ground state up to $O(\epsilon^2)$.  So, for $F_a$ to obey the error property, we need 
$\sum_k d_k n_j \psi_k+\mu+\epsilon n_j \nu$ to obey the error property.  Further, $\epsilon n_j \nu$ obeys the error property for any choice of $\nu$.
So, it remains to pick $\mu$ so that $\sum_{k\neq j} d_k n_j \psi_k+\mu$ obeys the error property.
Indeed, this can be done straightforwardly.  Let us write a basis for states where $|0\rangle$ is the state with no particles, and $|i,j,k,l\rangle= 
\psi^\dagger_i \psi^\dagger_j \psi^\dagger_k \psi^\dagger_l |0\rangle$.
The ground state has amplitude $1-O(\epsilon^2)$ on state $|0\rangle$, and it has amplitude $-\epsilon \frac{V_{ijkl}}{e_{i,j,k,l}}+O(\epsilon^2)$ on $|i,j,k,\rangle$.
Acting on $|i,j,k,l\rangle$ with $- d_k n_j \psi_k$ gives $\epsilon d_k  \frac{V_{ijkl}}{e_{i,j,k,l}} |i,j,l\rangle$, where $|i,j,l\rangle= \psi^\dagger_i \psi^\dagger_j \psi^\dagger_l |0\rangle$.
Thus, we may pick $\nu=\epsilon \sum_{k \neq j} \sum_{l} d_k  \frac{V_{ijkl}}{e_{i,j,k,l}} \psi^\dagger_i \psi^\dagger_j \psi^\dagger_k$ to obtain the error property.

Given such a decomposition of $H$, we now see how to choose things to appropriately reproduce $H_0$.
Let us ignore all terms of order $\epsilon$.  Let us now explicitly write the dependence on $a$ in $i,j,d_k,\ldots$ by adding $a$ in parentheses after each of them.
Then, we have
\begin{align}
\label{sumH0}
\sum_a c_a |F_a|^2=\sum_a c_a n_{i(a)} +\sum_a (c_a n_j \sum_{k\neq j(a)} d_k\psi^\dagger_{i(a)} \psi_k+\hc)
+\sum_a c_a n_j \bigl |\sum_{k\neq j(a)} d_k(a) \psi_k \bigr |^2+O(\epsilon).
\end{align}
So, to achieve the correct $H_0$ we need, first, $$\sum_{a,i(a)=i} c_a = e_i$$
for all $i$.  Let us ignore, for the moment, the third sum on the right-hand side of \cref{sumH0}.
The second sum on the right-hand side of \cref{sumH0} gives density-density terms and density-dependent hopping terms.  Indeed, we may adjust $j(a), c_a, d_k(a)$ to obtain \emph{any} density-density term or density-dependent hopping term from this second sum.
However, the third sum \emph{also} gives density-density terms and density-dependent hopping terms.  Since the second sum is $O(d)$ while the third sum is $O(d^2)$, for small enough $d$, the third sum becomes smaller than the second sum.
Thus, it is possible to adjust the $j(a),c_a,d_k(a)$ to obtain \emph{any sufficiently small} density-density terms and density-dependent hopping terms.
We emphasize that ``sufficiently small" does not mean that they are $O(\epsilon)$; rather, they are of order unity, but just sufficiently small compared to $1$.
In \cref{rangeofH}, we consider the set of density-density terms which can be decomposed in this way.

Let us now show that we have the closeness property, under one mild non-degeneracy assumption given later.  The terms $\sum_a c_a ( |\theta_a^\dagger|^2)$ are $O(\epsilon^2)$ so they do not affect the closeness property.  Similarly, any terms $\sum_a c_a ( |\theta_a|^2)$ arising from expanding $|F_a|^2$ are $O(\epsilon^2)$.  So, we may consider only linear order in $\theta$.

Now, the terms $\tau_i$ and $\mu$ in the definition  of $F_a$ are not things that we have a freedom to choose.  The term $\tau_i$ is fixed from our choice of  \cref{taudef},\cref{newdenom} and $\mu$ is fixed above.  However, we claim that, once we have adjusted the $j(a),c_a,d_k(a)$ to obtain the correct $H_0$, we will obtain the correct term $V$ in \cref{intformnew}, up to error $O(\epsilon^2)$.
This can be checked explicitly, but we can also use the same trick as used before in \cref{possapp}.  First, check that for all terms in $V$ which involve both creation and annihilation we get the correct result: this follows since we use the same such operators in $\tau_i$ as before and we have $\sum_{a,i(a)=i} c_a = e_i$ for all $i$.  That is, each $a$ with $i(a)=i$ contributes something towards $e_i$ and also contributes something to the terms in $V$ which involve both creation and annihilation operators and which include either $\psi_i$ or $\psi_i^\dagger$; since we correctly add up the terms to produce the correct $e_i$, we also produce the correct terms in $V$.  Next, all terms in $V$ which involve four creation or four annihilation operators must be correct up to $O(\epsilon^2)$ since we satisfy the error property; that is, since this decomposition of the Hamiltonian $H_{eff}$ obeys the error property in the ground state of $H$, we find that the expectation value of $H_{eff}$ in the ground state of $H$ is equal to the ground state energy of $H_{eff}$ up to $O(\epsilon^4)$, and hence, since both Hamiltonians are gapped for small $\epsilon$, the ground state must be equal up to $O(\epsilon^2)$, and hence the terms involving four creation or four annihilation operators must be correct up to $O(\epsilon^2)$.

The terms linear in $\nu(a)$ give
$\epsilon (\psi^\dagger_{i(a)}+\sum_{k\neq j} d_k \psi^\dagger_k) n_{j(a)} \nu(a)+\hc$.
Under a mild non-degeneracy assumption, 
by choosing $\nu(a)$, we can produce arbitrary $\sum_{i} \epsilon n_i V_i$ with these terms, so we achieve closeness for these terms.
The mild assumption is simplest to explain if $d_k=0$ for all $k\neq i(a)$.  In this case, we need that $d_{i(a)}\neq -1$, as if $d_{i(a)}=-1$ we would have
$\psi^\dagger_{i(a)}+\sum_{k\neq j} d_k \psi^\dagger_k=0$.  In the more general case, we need that the obvious linear equation defining $\mu$ given $\sum_{i} \epsilon n_i V_i$ has a solution, which is generically true.

Note, further, that it was necessary to include $\nu(a)$.  We already produce terms $\tau_i^\dagger n_j  \sum_{k\neq j} d_k \psi_k$ even if $\nu=0$, so we already produce $\epsilon n_i V_i$ terms.  So, $\nu(a)$ is included as a way to cancel these terms.

Finally, we claim that this obeys the closure property.
Indeed, because we have added in the term $|\theta_a|^2$, 
all terms which are quadratic in $\theta,\theta^\dagger$ appear as an anti-commutator and hence reduce to quartic terms in $\psi,\psi^\dagger$, possibly multiplied by a number operator.  

\subsection{Example of How It Can Work}
It may be somewhat surprising that this works, given the flexibility in choosing different $c,d$.  That is, how do all these possible choices still lead to a decomposition of the same Hamiltonian?  So
let us now give a simple example of how this can work.
Consider a Hamiltonian with four fermionic modes given by
$$
H=\sum_{i=1}^4 n_i + U n_1 n_2 + (\epsilon \psi^\dagger_1 \psi^\dagger_2 \psi^\dagger_3 \psi^\dagger_4 + \hc).
$$
Consider the sum-of-squares
\begin{align}
\nonumber
|\psi_1+d_1 n_2 \psi_1+\epsilon\frac{1+d_1}{4+U}\psi^\dagger_2\psi^\dagger_3 \psi^\dagger_4 |^2 
\\+| \psi_2+d_2 n_1 \psi_2 -\epsilon\frac{1+d_2}{4+U} \psi^\dagger_1\psi^\dagger_3 \psi^\dagger_4  |^2
\\+|\psi_3 +\frac{\epsilon}{4+U} \psi^\dagger_1 \psi^\dagger_2 \psi^\dagger_4|^2
\\+|\psi_4 - \frac{\epsilon}{4+U} \psi^\dagger_1 \psi^\dagger_2 \psi^\dagger_3|^2.
\end{align}
Here, the different $d_1,d_2$ correspond to choices of different $d$ in our decomposition above.
To obtain the correct $U$ we need
\be
\label{rightU}
2d_1+d_1^2+2d_2+d_2^2=U.
\ee
The coefficient of the term $\psi^\dagger_1 \psi^\dagger_2 \psi^\dagger_3 \psi^\dagger_4$ in this sum-of-squares is
$$\epsilon \frac{(1+d_1)^2+(1+d_2)^2+2}{4+U}
+O(\epsilon^2).$$
By \cref{rightU}, $(1+d_1)^2+(1+d_2)^2=2+U$, so the coefficient is indeed equal to $\epsilon+O(\epsilon^2)$ as required for any choice of $d_1,d_2$ obeying \cref{rightU}.

\section{Range of Hamiltonians Covered by the Decomposition}
\label{rangeofH}
Now consider the possible $H_0$ which can be constructed with this decomposition.  
We take $H=H_0$, ignoring all terms of order $\epsilon$.
Further, we assume that $H_0=\sum_{i} e_i n_i + \sum_{i<j} U_{i,j} n_i n_j.$

Each term in the sum-of-squares will be of the form
$c_a |\psi_i + d_a n_j  \psi|^2=c_a n_i + c_a(2d_a+d_a^2)n_i n_j$.
The sum of $c_a$ over all $a$ with $i(a)=i$ is equal to $e_i$ and each $c_a$ is non-negative.  However, by taking $d_a$ and $c_a$ small in the the limit we also include terms $|\psi_i n_j|^2=n_i n_j$.  Let us include this limiting case, allowing us to produce all terms $U_{i,j} n_i n_j$ with $U_{i,j}>0$ in this way.
This reduces the problem of whether we can construct an arbitrary $H_0$ to the problem of whether we can construct an $H_0$ with all $U_{i,j}\leq 0$.

To produce a term $U_{i,j} n_i n_j$ with $U_{i,j}<0$, the most efficient way (producing the most negative $n_i n_j$ for the smallest total coefficient in front of $n_i$ and $n_j$) is take $d=-1$, giving either $n_i-n_i n_j$ or $n_j - n_i n_j$, and by taking linear combinations of this we can produce any $a n_i + b n_j + u n_i n_j$ so long as $u\geq -a -b$ with $a,b\geq 0$.

Thus, a given $H_0$ of this form that can be constructed in this way if and only if, for each $i<j$ with $U_{i,j}<0$, we can write $U_{i,j}$ as minus the sum of two different ``weights", $$U_{i,j}=-w_i(i,j)-w_j(i,j),$$ with both $w_i(i,j),w_j(i,j)\geq 0$, such that $\sum_j w_i(i,j) \leq e_i$ for all $i$.
This is equivalent to a network flow problem.  We have a source vertex.  We have a set $P$ of vertices labelled by pairs $i,j$ with $i<j$ for each $i,j$ with $U_{i,j}<0$, with an edge from the source vertex to each vertex in the set $P$, with capacity $-U_{i,j}$ on each such edge.  We have another set $Q$ of vertices labelled by an index $i$ and each vertex in $P$ labelled by a pair $i,j$ is connected to both $i$ and $j$, by an edge with unlimited capacity.  Finally, we have a sink vertex and each vertex in $Q$ labelled by index $i$ is connected to the sink vertex with capacity $e_i$.  The given $H_0$ can be produced if a flow exists with capacity equal to the sum of $-U_{i,j}$ over all $i,j$ such that $U_{i,j}<0$.

The equivalence to a network flow problem lets us show that \emph{an $H_0$ with all $U_{i,j}\leq 0$ can be constructed if and only if a ground state of such an $H_0$ has all $n_i=0$}.  (We say ``a" ground state as multiple states may have the same energy.)  
Remark: this does not mean that an $H_0$ with some $U_{i,j}>0$ can be constructed if it has a ground state with all $n_i=0$; instead, one must take the $H_0$ with some $U_{i,j}>0$, set the positive $U_{i,j}$ to zero to obtain some other $H_0$, and see if that other $H_0$ has a ground state with all $n_i=0$.

To prove this claim,
suppose not, so we have some example where a ground state has all $n_i=0$ and where all $U_{i,j}\leq 0$, but the given flow does not exist.  Then, by max flow-min cut, there exists some cut $C$ with capacity less than $-\sum_{i<j} U_{i,j}$.  Since the edges from vertices in $P$ labelled by pairs $i,j$ to single indices $i$ or $j$ have unlimited capacity, the cut does not include these edges.  Suppose the cut contains some edge from the source to a vertex labelled by a pair $i,j$.  We can consider a new $H_0$, called $H_0'$, where this particular $U_{i,j}$ is set to zero, with all other terms left unchanged.  Let $U'_{i,j}$ be the coefficient of the $n_i n_j$ term in this new $H_0'$.  Then, $H_0'$ has some cut $C'$ with capacity less than $-\sum_{i<j} U'_{i,j}$, as can be seen by taking $C'$ to include all edges in $C$, except the edge from source to $i,j$.  Further, since $U_{i,j}<0$, if some ground state of $H_0$ had all $n_i=0$, then the same holds for $H_0'$ as adding $-U_{i,j} n_i n_j$ to $H_0$ can only increase the energy of an eigenstate.  Hence, we may reduce to the case where no edges in the cut go from the source to a vertex labelled by a pair $i,j$.  Then, the edges in the cut go from vertices labelled by a single index $i$ to the sink, and hence they define some set $S$ of indices $i$.  
So, summing the capacity of the edges in the cut, we have $\sum_{i} e_i < -\sum_{i<j} U_{i,j}$
The nonvanishing $U_{i,j}$ must only be such that $i,j\in S$; if not, we do not have a cut.  So the fact that
 $\sum_{i} e_i < -\sum_{i<j} U_{i,j}$ means that the state with $n_i=1$ for $i\in S$ and $n_i=0$ otherwise has energy $<0$, giving a contradiction.

\section{Extensions}
We have given a decomposition that gives the energy to $O(\epsilon^4)$ even when there are density-density and density-dependent hopping terms which are order unity rather than order $\epsilon$.
Let us discuss some extensions and applications.

First, we emphasize that the density-dependent hopping terms we produce are of the form 
$n_i (\psi^\dagger_j \psi_k + \hc)$, rather than $n_i (\psi^\dagger_j \psi^\dagger_k+\hc)$, and this is after particle-hole conjugation is done.  So, any physical density-dependent hopping terms must be of the form where $j,k$ are both occupied or both empty in the ground state of $H_0$.

Now, since density-dependent hopping terms are not, in practice, as strong as density-density interaction terms, we hope that this will not be a significant limitation in practice; the density-dependent hopping terms which are of the form $n_i (\psi^\dagger_j \psi^\dagger_k+\hc)$ can be incorporated into $V$ where they are treated as terms of order $\epsilon$.  The reason that we have this limitation on allowed density-dependent hopping is as follows.
Suppose, to include these terms, we have taken 
$F_a=\psi_i + n_j  \sum_{k\neq j} d_k \psi_k+\sum_{k\neq j} d'_k \psi^\dagger_k+\theta_a,$ with some scalar $d'_k$.
Then, $F_a$ acting on the ground state would have some amplitude proportional to $\epsilon$ on states with \emph{five} excitations, and so $\mu$ would need to include some $O(\epsilon)$ term with five creation operators.
This could be done, but would make the construction more complicated, as the closure property would become more difficult.

A second extension is obtaining a spin-spin interaction term.  As we have explained, this is simply a sum of different density-density interaction terms in different single-particle bases.  We can still use the same decomposition of the form $H=\sum_a c_a ( |F_a|^2+|\theta_a^\dagger|^2)$, but now for each $F_a$ we may choose a different single-particle basis, with different bases related by spin-rotation, summing density-density terms over different bases to obtain the spin-spin interaction.

Another possible extension may be to allow the use of more than one number operator in each $F_a$.  For example, we might pick
$F_a=\psi_i + n_j  \sum_{k\neq j} d_k \psi_k+ n_{j'} \sum_{k\neq j'} d'_k \psi_k+
\theta_a,$
for some $j' \neq i,j$.  In this case, however, we start to generate terms $n_i V_i$ and also $n_i n_j V_{i,j}$ in the effective Hamiltonian, so $\theta_a$ likely needs to be of the form
$\theta_a=\tau_i + \mu+\epsilon n_j \nu+\epsilon n_{j'} \nu' + \epsilon n_j n_{j'} \nu''.$  This extension would be numerically more complicated but may increase the range of possible Hamiltonians that can be decomposed in this way.

Finally, in \cite{hastings2024improving}, a generalized algorithm was given.  The decomposition of \cref{Hdecomp} is one possible decomposition.
It is obtained by finding such a decomposition of the trial Hamiltonian $H'$ and then adjusting the trial Hamiltonian.  In the generalized algorithm, we regard this decomposition of the trial Hamiltonian as defining some quadratic in the operators $\psi_i^\dagger,\tau_i^\dagger,\psi_i,\tau_i$.  We then pretend that the anti-commutator $\{\tau_i^\dagger,\tau_j\}$ is equal to its expectation value in the no-particle state $|0\rangle$.  Using this approximation, we then find an optimal decomposition of this given quadratic as a sum-of-squares.  In such an optimal decomposition, the term $|\tau^\dagger|^2$ may be replaced by something similar to $|\tau^\dagger + \epsilon \psi^\dagger |^2$.
Here, ``optimal" means that it would provide the best bound on the energy.  We then use this decomposition to find the effective Hamiltonian.  (To emphasize, our ``pretense" about the value of the anti-commutator is simply used to compute the decomposition, while when we compute the effective Hamiltonian we use the true value of the anti-commutator).  This method is completely analogous to the following: suppose $\tau^\dagger,\tau$ really were fermion creation and annihilation operators on some other fermion mode, rather than being cubics, and consider the Hamiltonian $\psi^\dagger \psi + \epsilon (\psi^\dagger\tau+\hc),$ which is a simple non-interacting fermionic Hamiltonian.  We could decompose this as $|\psi+\epsilon \tau|^2 + \epsilon^2 |\tau^\dagger|^2-\epsilon^2$, which would let us lower bound the ground-state energy by $-\epsilon^2$ which is correct to order $\epsilon^2$ but not correct to order $\epsilon^4$.  Instead a better decomposition can be given, where the terms in the sum-of-squares are squares of eigenoperators of the Hamiltonian, and this is exact to all orders in this case.  Applying this generalized algorithm when $\tau$ is a cubic, the energy bound can improve; it will still be only accurate to $O(\epsilon^4)$ in general, but the error at that order may decrease.   A similar generalized algorithm here might mean regarding 
the decomposition as some quadratic in 
$\psi_i + n_j  \sum_{k\neq j} d_k \psi_k$ and
$\theta_a=\tau_i + \mu+\epsilon n_j \nu.$  One might more generally regard it as a quadratic in $\psi_i, n_j  \sum_{k\neq j} d_k \psi_k,\theta_i,\mu,\epsilon n_j \nu$, but this would complicate things.  We leave this generalized algorithm to a future implementation of the method here.
\bibliography{ninj-ref}
\end{document}